\documentclass[twocolumn,prb,showpacs,multicol,amsmath,amssymb]{revtex4}
\usepackage[dvips]{graphicx}
\usepackage{subfigure}
\usepackage{graphicx}
\usepackage{dcolumn}
\usepackage{bm}
\usepackage{graphics}
\usepackage{epsfig,color}

\newcommand{\be}{\begin{equation}}
\newcommand{\ee}{\end{equation}}
\newcommand{\bea}{\begin{eqnarray}}
\newcommand{\eea}{\end{eqnarray}}

\begin{document}
\title[M. R. Soltani ]{Quantum correlations in the 1D spin-1/2 Ising model with added Dzyaloshinskii-Moriya interaction}
\author{M. R. Soltani$^{1}$\footnote{email: m.r.soltani.em@gmail.com\\ Tel: (+98) 912 1768682\\Fax: (+98)21 44033189}, J. Vahedi$^{2}$, S. Mahdavifar$^{3}$}
\address{ $^{1}$ Department of Physics, College of Science,Yadegar-e-Imam Khomeini(RAH)Branch, Islamic Azad University, Tehran, Iran.\\
$^{2}$ Department of Physics, Sari Branch, Islamic Azad University, Sari, Iran.\\
$^{3}$Department of Physics, University of Guilan, 41335-1914, Rasht, Iran.}
\date{\today}

\begin{abstract}
We have considered the 1D spin-1/2 Ising model with added Dzyaloshinskii-Moriya (DM) interaction and presence of a
uniform magnetic field. Using the mean-field fermionization approach the energy spectrum in an infinite chain is obtained.
The quantum discord (QD) and concurrence between nearest neighbor (NN) spins at finite temperature are specified as a
function of mean-field order parameters. A comparison between concurrence and QD is done and differences are obtained.
The macroscopic thermodynamical witness is also used to detect  quantum entanglement region in solids within our model.
We believe  our results are useful in the field of the quantum information processing.
\pacs{03.65.Ud; 03.67.Mn; 75.10.Pq}
\end{abstract}

\maketitle

\section{INTRODUCTION}\label{sec1}
In the last few years, the entanglement which comes from quantum information may lead to the further
insight into the other areas of physics such as condensed matter and statistical mechanics \cite{a1,a2,a3,a4,a5,a6,a7,a8}.
Entanglement can be of various types, e.g., bipartite, multipartite, entanglement entropy etc,
for which a number of quantitative measures exist. The entanglement content of quantum states and its
variations have been extensively investigated in recent years\cite{a2,a3,a4,a5}.

However, It is not the whole story and the entanglement dose not show all quantum correlations
among different constituents of a quantum system.
The notion of quantum discord (QD) introduced by Ollivier and Zurek\cite{a9} can
measure the quantumness of correlations (including entanglement).
The QD is considered as quantum correlation and  quantum resource
likewise entanglement. It is curious to ask what the characteristic of QD
is at finite temperature, and what the differences are between thermal QD ($TQD$) 	
and entanglement in a system. The authors pointed out that the QD, in
contrast to the entanglement of formation (EOF), can detect
those quantum correlations present in certain separable mixed
states\cite{a9}.
\par
In the topic of quantum magnets, a lack of inversion symmetry make ones, to consider an extra exchange
 coupling between spins in the magnetic materials than the usual and well
known the isotropic exchange $\vec{S}_{i}\cdot\vec{S}_{j}$. In this respect,
Dzyaloshinskii has shwon\cite{Dzyaloshinskii58} that an antisymmetric exchange
$\vec{D}_{ij}\cdot(\vec{S}_{i}\times\vec{S}_{j})$ should be considered in these magnetic materials.
Later, Moriya has shown\cite{Moriya60} that inclusion of the spin orbit coupling on the magnetic ions in the 1st and 2nd order
leads to antisymmetric and anisotropic exchange respectively.

This interaction is, however, rather difficult to handle analytically,
but it is one of the agents responsible for magnetic frustration. Since this interaction may
induce spiral spin arrangements in the ground state\cite{sa6}, it
is closely involved with ferroelectricity in multiferroic spin
chains\cite{sa7,sa8}. Besides, the DM interaction plays an important
role in explaining the electron spin resonance experiments
in some one-dimensional antiferromagnets\cite{sa9}. Moreover,
the DM interaction modifies the dynamic properties\cite{sa10}
and the quantum entanglement\cite{sa11,sa111} of spin chains\cite{sa12}.
Behavior of quantum and classical correlations in the XY-spin chain with DM
interaction also was discussed\cite{sa13}. Using Bethe Ansatz technique, authors in Ref.\cite{sa14},
have investigated the QD in the spin-1/2 XXZ chain system.
\par
In this article we study the thermal behavior of the pairwise QD and EOF, in an infinite  spin-$\frac{1}{2}$ Ising chain
with added Dzyaloshinskii-Moriya interaction subjected to an external longitudinal
magnetic field. Such contributions, to the best knowledge of authors, have not been considered
in previous works. Here, using the Jordan-Wigner transformations\cite{sa15} we diagonalize the
Hamiltonian of the system and then obtain the QD and EOF as a function of order parameters.
We also use macroscopic thermodynamic properties as entanglement witness to detect the presence of entanglement in solid state systems.

This paper is organized as follows. In Sect. 2 we briefly review the definition of the QD and the EOF from the
 information theory point of view. In Sect. 3 we introduce the  spin-1/2 Ising chain with DM interaction and describe
briefly the technique of Jordan-Wigner transformation and the mean-filed approximation used to diagonalize the Hamiltonian.
In Sect. 4 we analyze the behavior of the QD and the EOF  under the influence of the DM interaction, the longitudinal magnetic field and the temperature.
In Sect. 5 we use macroscopic thermodynamic witness to detect  quantum entanglement region in solids which is described by our model.
Finally, in Sect. 6 a brief summary is given.

\section{QUANTUM CORRELATION}\label{sec2}
\subsection{Quantum Discord}

We first present a brief review of the QD,
which is defined as the difference between two expressions of
mutual information extended from the classical to the quantum
system.
In classical information theory, the total correlation
between two systems $A$ and $B$, whose state is mathematically
represented by a joint probability distribution $p(A,B)$, can be
obtained by the mutual information

We first present a brief review of the QD,
which is defined as the difference between two expressions of
mutual information extended from the classical to the quantum
system.
In classical information theory, the total correlation
between two systems $A$ and $B$, whose state is mathematically
represented by a joint probability distribution $p(A,B)$, can be
obtained by the mutual information

\begin{eqnarray}
\mathcal{I}(A:B)=H(A)+H(B)-H(A,B),
\label{eq1}
\end{eqnarray}
where $H(\cdot) = -\sum_i p_i \log_2 p_i$ is the Shannon entropy,
and $p_i$ shows the probability of an event $i$ relevant to the system $A$ or $B$ or the joint system $AB$. By using the Bayes
rule, one can rewrite the mutual information as
\begin{eqnarray}
\mathcal{I}(A:B)=H(A)-H(A|B),
\label{eq2}
\end{eqnarray}
where $H(A|B)=H(A,B)-H(B)$ denotes the classical conditional
entropy and it is employed to quantify the ignorance
(on average) regarding the value of $A$ when one knows $B$. In
classical information theory, these two expressions are equivalent,
but there is a difference between them in the
quantum world. In order to generalize these expressions to the quantum world,
we replace classical probability distribution
by the density operator $\rho$ and the Shannon entropy by the von
Neumann entropy $S(\rho)=-Tr(\rho\log_2\rho)$. In particular, if $\rho_{AB}$
denotes the density operator of a composite bipartite system
$AB$, then $\rho_A(\rho_B)$ the reduced density matrix of subsystem $A(B)$.
Now one can give the quantum versions of Eqs.(\ref{eq1}) and (\ref{eq2}),
respectively:
\begin{eqnarray}
\mathcal{I}(\rho_A:\rho_B)&=&S(\rho_A)+S(\rho_B)-S(\rho_{AB}),\\
\mathcal{L}(\rho_A:\rho_B)&=&S(\rho_A)-S(\rho_A|\rho_B),
\label{eq3}
\end{eqnarray}
where $S(\rho_A|\rho_B)$ is a quantum generalization of the conditional
entropy for $A$ and $B$, and it cannot be directly obtained
via the replacement of the Shannon entropy by the von Neumann
entropy. To get access  quantum conditional entropy, we choose
projective measurements on $B$ described by a complete set
of orthogonal projectors, $\Pi_i$, corresponding to outcomes
labeled by $i$. Once the measurement is made, the state of the system is given by
\begin{eqnarray}
\rho_i=\frac{1}{p_i}\left(I\otimes\Pi_i^B\right)\rho_{AB}\left(I\otimes\Pi_i^B\right),
\label{eq4}
\end{eqnarray}
with
\begin{eqnarray}
p_i=Tr\left(\left(I\otimes\Pi_i^B\right)\rho_{AB}\left(I\otimes\Pi_i^B\right)\right),
\label{eq5}
\end{eqnarray}
where $I$ is the identity operator for the subsystem $A$ and $p_i$ denotes the probability
for obtaining the outcome $i$. The quantum analogue of the conditional entropy is given by
\begin{eqnarray}
S(\rho_{AB}|\{\Pi_i^B\})=\sum_ip_iS(\rho_i),
\label{eq6}
\end{eqnarray}
and then the quantum extension of classical mutual information is given by
\begin{eqnarray}
\mathcal{L}(\rho_{AB}|\{\Pi_i^B\})=S(\rho_A)-S(\rho_{AB}|\{\Pi_i^B\}).
\label{eq7}
\end{eqnarray}
When projective measurements are made on the subsystem $B$, the non-classical
correlations between the subsystems are removed. Since the value of
$\mathcal{L}(\rho_{AB}|\{\Pi_i^B\})$ depends on the choice of $\{\Pi_i^B\}$,
$\mathcal{L}$ should be maximized over all $\{\Pi_i^B\}$ to ensure that
it contains the whole of the classical correlations. Thus the quantity
\begin{eqnarray}
\mathcal{C(}\rho_{AB})=\max_{\{\Pi_i^B\}} \left(\mathcal{L}(\rho_{AB}|\{\Pi_i^B\})\right),
\label{eq8}
\end{eqnarray}
provides a quantitative measure of the total classical correlations\cite{sa16,sa17}.
Once $\mathcal{C}$ is in hand, QD can be obtained by subtracting it from
the quantum mutual information
\begin{eqnarray}
\mathcal{Q}(\rho_{AB})=\mathcal{I}(\rho_A:\rho_B)-\mathcal{C(}\rho_{AB}).
\label{eq9}
\end{eqnarray}
\subsection{Entanglement of formation}
Here, we give a short introduction about  the entanglement of formation, a well known tool  for measuring the entanglement\cite{sa18}. To this end, we needed just to find the reduced density matrix $\rho_{i,j}$. It has been shown by Wootters\cite{sa18}  that  for a pair of qubits, that the concurrence $C$, which changing from $0$ to $1$, can be served as a good measure of entanglement. Now the concurrence between two  sites  is defined as\cite{sa18}
\begin{equation}
C(\rho)=max\{0,\lambda_1-\lambda_2-\lambda_3-\lambda_4\},
\end{equation}
where the $\lambda_i$'s are the eigenvalues of  $R\equiv\sqrt{\sqrt{\rho}\tilde{\rho}\sqrt{\rho}}$
 which   $\tilde{\rho}=(\sigma^y\otimes\sigma^y)\rho^\ast(\sigma^y\otimes\sigma^y)$  is the spin-flipped state
is and $\sigma$ is the Pauli matrix . Now we can write the entanglement bewteen a pair of qubite as
 \begin{equation}
E(\rho)=ho)=g\left(\frac{1-\sqrt{1-C^2}}{2} \right),
\end{equation}
where $g$ is the binary entropy function
\begin{equation}
g(x)=-x\log_2x-(1-x)\log_2(1-x).
\end{equation}
Now, the entanglement of formation is given in terms of the concurrence $C$.

\section{The model} \label{sec3}
We consider the 1D spin-1/2 Ising model with added Dzyaloshinskii-Moriya interaction
subjected to an external magnetic field  in the $z$ direction (LF).
The Hamiltonian of the model is written as
\begin{eqnarray}
H&=&J\sum_{n}S^{z}_{n}S^{z}_{n+1}+\sum_{n}\mathbf{D}\cdot \left(\mathbf{S}_{n}\times \mathbf{S}_{n+1}\right) \nonumber\\
&-&h\sum_{n}S^{z}_{n},
\label{eq15}
\end{eqnarray}
where $\mathbf{S}_{n}$ is the spin-1/2 operator on the $n$-th site, $h$ is the longitudinal
magnetic field (LF), $J>0$ is the antiferromagnetic coupling constant and $\mathbf{D}=D\hat{z}$
denotes a uniform DM vector. In the absence of the LF, the ground state phase diagram of
the model is known\cite{sa20, Jafari08, Soltani12}. The spectrum of the model for $-D< J \leq D$ is
gapless and the system is in the so-called Luttinger-liquid (LL)
phase with a power-law decay of correlations. In the Ising-like region $D<J$, the ground state of
the model has a Neel long-range order along the $Z$ axis
and there is a gap in the excitation spectrum. In the presence of the LF, the spectrum remains
gapless if the field does not exceed a saturated critical value $h_c=D+J$. At zero temperature,
in the absence  of the LF, nearest neighbor (NN) spins  are entangled and by increasing the LF
 the concurrence decreases and will be equal to zero at the critical LF, $h_{c}$.  In the saturated
 ferromagnetic phase, NN spins are not entangled\cite{Jafari08}.

Theoretically, the energy spectrum is needed to investigate the thermodynamic properties of the model.
In this respect, we implement the Jordan-Wigner transformation to fermionize
the Hamiltonian (Eq.~(\ref{eq15})). Using the Jordan-Wigner transformations
\begin{eqnarray}
S^{+}_{n}&=&a_{n}^{\dag}e^{i\pi\sum^{n-1}_{m=1}a^{\dag}_{m}a_{m}},\nonumber\\
S^{-}_{n}&=&e^{-i\pi\sum^{n-1}_{m=1}a^{\dag}_{m}a_{m}}a_{n},\nonumber\\
S^{z}_{n}&=&a_{n}^{\dag}a_{n}-\frac{1}{2},
\label{eq17}
\end{eqnarray}
Hamiltonian (\ref{eq15}) takes the form
\begin{eqnarray}
H_{F}&=&\sum_{n}\frac{-i D}{2}\left(a^{\dag}_{n}a_{n+1}+a_{n} a^{\dag}_{n+1}\right)\nonumber\\
     &+& J \sum_{n} \left(a^{\dag}_{n}a_{n}a^{\dag}_{n+1}a_{n+1}\right)\nonumber\\
      &-&(J+h)\sum_{n}a^{\dag}_{n}a_{n}+constant~.
\label{eq18}
\end{eqnarray}
Treating the Hamiltonian $H_{F}$ in the mean-field approximation, the
interacting fermionic system reduces to a 1D system of the non-interacting quasi particles\cite{Dmitriev02}
\begin{eqnarray}
H_{MF}&=&\frac{-i D}{2}\sum_{n}\left(a^{\dag}_{n}a_{n+1}+a_{n} a^{\dag}_{n+1}\right) \nonumber \\
      &-&J \gamma_3 \sum_{n}\left(a_{n}a_{n+1}+a^{\dag}_{n+1}a^{\dag}_{n}\right)\nonumber \\
      &-&J \gamma_2 \sum_{n}\left(a^{\dag}_{n+1}a_{n}+a^{\dag}_{n}a_{n+1}\right)\nonumber \\
      &-&(2 \gamma_1 J-J-h) \sum_{n} a^{\dag}_{n}a_{n},
\label{eq18-1}
\end{eqnarray}
where $\gamma_{i} (i=1 ,2 ,3)$ are the mean-field order parameters,
\begin{eqnarray}
\gamma_{1}&=&\langle a^{\dag}_{n}a_{n}\rangle,\nonumber\\
\gamma_{2}&=&\langle a^{\dag}_{n}a_{n+1}\rangle,\nonumber\\
\gamma_{3}&=&\langle a^{\dag}_{n}a_{n+1}^{\dag}\rangle.
\label{eq22}
\end{eqnarray}
By performing a Fourier transformation into the momentum space as
$a_{n} = \frac{1}{\sqrt{N}} \sum ^{N} _{n=1} e^{-ikn} a_{k}$, the mean field Hamiltonian takes the form
\begin{eqnarray}
H_{MF}&=&\sum_{k>0}a(k) \left(a^{\dag}_{k}a_{k}+a^{\dag}_{-k}a_{-k}\right) \nonumber \\
      &+& \sum_{k>0}b(k) \left(a^{\dag}_{-k}a_{-k}+a^{\dag}_{k}a_{k}\right) \nonumber \\
      &-&i \sum_{k>0}c(k) \left(a_{k}a_{-k}-a^{\dag}_{-k}a^{\dag}_{k}\right),
\label{eq18-2}
\end{eqnarray}
where
\begin{eqnarray}
a(k)&=& 2 J \gamma_{2} \cos(k)+2 J \gamma_{1}-J-h, \nonumber\\
b(k)&=&D \sin(k), \nonumber\\
c(k)&=&2 J \gamma_{3} \sin(k).
\label{eq19}
\end{eqnarray}
Finally, by using the following Bogoliubov transformation
\begin{eqnarray}
a_{k}=\cos(k) \beta_k+i \sin(k)\beta^{\dag}_{-k},
\label{u-t}
\end{eqnarray}
the diagonalized Hamiltonian is given by
\begin{eqnarray}
H_{MF}&=& \sum_{k} \varepsilon_k \beta^{\dag}_{k}\beta_{k},
\label{D-hamiltoni}
\end{eqnarray}
where $\varepsilon_k=\sqrt{a(k)^{2}+c(k)^{2}}-b(k)$ is the energy spectrum.
Since, the Hamiltonian
(\ref{eq15}) exhibits both translational and U(1) invariance, density matrix will be given by
\begin{equation}
\rho_{i,i+1}=
\begin{pmatrix}
\rho_{11} & 0 & 0 & 0\\
0 & \rho_{22} & \rho_{23} & 0\\
0 & \rho_{23} & \rho_{22} & 0\\
0 & 0 & 0 & \rho_{44}\\
\end{pmatrix},
\label{eq23}
\end{equation}
where
\begin{eqnarray}
\rho_{11}&=&\langle n_j n_{j+1}\rangle (n_j=a_{j}^{\dagger}a_{j})\rangle,\nonumber\\
\rho_{22}&=&\langle n_j(1-n_{j+1})\rangle,\nonumber\\
\rho_{33}&=&\langle n_{j+1}(1-n_{j})\rangle,\nonumber\\
\rho_{44}&=&\langle 1-n_j- n_{j+1}+n_j n_{j+1}\rangle,\nonumber\\
\rho_{23}&=&\langle a_{j}^{\dagger}a_{j+1}, \rangle.
\label{eq24}
\end{eqnarray}
and
\begin{eqnarray}
\langle n_j \rangle&=&\gamma_{1}\nonumber \\
&=&\frac{1}{2}+\frac{1}{2 \pi} \int_{-\pi}^{\pi} \frac{a(k)}{\varepsilon(k)}(\frac{1}{1+e^{\beta \varepsilon(k)}}-\frac{1}{2}) dk~,
\end{eqnarray}
\begin{eqnarray}
\rho_{23}&=&\gamma_{2} \\
&=&\frac{1}{2 \pi} \int_{-\pi}^{\pi} \cos(k) (\frac{a(k)}{\varepsilon(k)}) (\frac{1}{1+e^{\beta \varepsilon(k)}}-1/2) dk~, \nonumber
\end{eqnarray}
where $\beta=\frac{1}{k_{B} T}$ and the Boltzmann constant is considered equal to one ($k_B=1$). One should note that the Fermi distribution function is $f(k)=\frac{1}{1+e^{\beta \varepsilon(k)}}$. Using the solution of the retarded Green's function\cite{Fetter71},  $\rho_{11}$ approximately is obtained as $\langle n_j n_{j+1}\rangle=\gamma_{1}^{2}-\gamma_{2}^{2}$. Thus the concurrence as a function of the mean-field order parameters is given as

\begin{eqnarray}
C_{j}&=&max\{0,2 (|\rho_{23}|-\sqrt{\rho_{11} \rho_{44}})\},\nonumber \\
&=&max\{0,2 (|\gamma_{2}|-\sqrt{(\gamma_{1}^2-\gamma_{2}^2)[(1-\gamma_{1})^{2}-\gamma_{2}^2]}.\nonumber \\
\label{concurrence}
\end{eqnarray}
 At zero temperature, the magnetization in the absence of the LF is zero, therfore $\gamma_1=1/2$.
 In addition, there is no spin correlation in the $xy$ plane and $\gamma_2=0$. This leads to
  unentangled state in the pure Ising model. On the other hand, in the pure DM interaction model,
   there is not any interaction along the $z$ direction and magnetization is zero ($\gamma_1=1/2$). In this case concurrence will be equal to $max\{0,2 (|\gamma_{2}|-|\frac{1}{4}-\gamma_{2}|$\}.
    One can show that the mean-field order parameter $\gamma_{2}$ only in the RVB state takes
    the maximum value $\frac{1}{4}$. Thus, the concurrence gets the maximum value, $C_{j}^{max}=0.5$.
    Due to the induced quantum fluctuations in the 1D spin-1/2 Ising model with added DM interaction,
    the concurrence should be less than $0.5$.
Now for studying the QD we follow the prescription which is introduced by M. Sarandy\cite{sa22}.
By defining new variables
\begin{eqnarray}
c_1&=&2\rho_{23}=2 \gamma_2,\nonumber\\
c_2&=&\rho_{11}+\rho_{44}-2\rho_{22}=1-4 \gamma_1+ 4(\gamma_{1}^{2}-\gamma_{2}^{2}),\nonumber\\
c_3&=&\rho_{11}-\rho_{44}=2 \gamma_1-1~.
\label{eq26}
\end{eqnarray}
The eigenvalues of $\rho_{i,i+1}$ can be read as
\begin{eqnarray}
\lambda_0&=&\gamma_1^{2}-\gamma_2^{2},\nonumber\\
\lambda_1&=&1+\gamma_1^{2}-\gamma_2^{2}-2 \gamma_1,\nonumber\\
\lambda_2&=&-\gamma_1^{2}+\gamma_2^{2}+\gamma_1+\gamma_2,\nonumber\\
\lambda_3&=&-\gamma_1^{2}+\gamma_2^{2}+\gamma_1-\gamma_2.
\label{eq27}
\end{eqnarray}
Therefore, the mutual information is given by
\begin{eqnarray}
\mathcal{I}(\rho_{i,i+1})=S(\rho_i)+S(\rho_{i+1})+\sum_{\alpha=0}^{3}\lambda_{\alpha}\log\lambda_{\alpha},
\label{eq28}
\end{eqnarray}
where
\begin{eqnarray}
S(\rho_i)&=&S(\rho_{i+1})=\\
&-&\Big[\gamma_1\log(\gamma_1)+(1-\gamma_1)\log(1-\gamma_1) \Big].\nonumber
\label{eq29}
\end{eqnarray}
Classical correlations is obtained by following procedure.
We first introduce a set of projectors for a local measurement on the part $(i+1)=B$ given by
$\{B_k=V\Pi_kV^\dag\}$ where $\{\Pi_k=|k\rangle \langle k|:k=0,1\}$ is the set of projectors
on the computational basis $|0\rangle\equiv|\uparrow\rangle$ and $|1\rangle\equiv|\downarrow\rangle$
and $V \in U(2)$. Note that the projectors $B_k$ represent  an arbitrary local measurement on $B$.
We parameterize V as
\begin{equation}
V=
\begin{pmatrix}
\cos \frac{\theta}{2} & \sin \frac{\theta}{2}e^{-i\phi} \\
\sin \frac{\theta}{2}e^{i\phi} & -\cos \frac{\theta}{2} \\
\end{pmatrix},
\label{eq30}
\end{equation}
where $0\leq\theta\leq\pi$ and $0\leq\phi<2\pi$. Note that $\theta$ and
$\phi$ can be interpreted as the azimuthal and polar angles,
respectively, of a qubit over the Bloch sphere. By using Eq.~(\ref{eq5}),
One can show that the state of the system after measurement ${B_k}$ will change
to one of the states
\begin{eqnarray}
\rho_0=\left(\frac{I}{2}+\sum_{j=1}^{3}q_{0j}S^{j}\right)\otimes(V\Pi_0V^\dag),
\label{eq32}
\end{eqnarray}
\begin{eqnarray}
\rho_1=\left(\frac{I}{2}+\sum_{j=1}^{3}q_{1j}S^{j}\right)\otimes(V\Pi_1V^\dag),
\label{eq33}
\end{eqnarray}
where
\begin{eqnarray}
q_{k1}&=&(-1)^kc_1\left[\frac{\sin\theta\cos\phi}{1+(-1)^kc_3\cos\theta}\right],\nonumber\\
q_{k2}&=&\tan\phi q_{k1},\nonumber\\
q_{k3}&=&(-1)^k\left[\frac{c_2\cos\theta+(-1)^kc_3}{1+(-1)^kc_3\cos\theta}\right].
\label{eq34}
\end{eqnarray}
Then, by evaluating the von Neumann entropy from
Eqs.~(\ref{eq32}) and (\ref{eq33}) and using $S(V\Pi_0V^\dag)=0$, we obtain
\begin{eqnarray}
S(\rho_k)=-(\frac{1+\theta_k}{2})\log(\frac{1+\theta_k}{2})+(\frac{1-\theta_k}{2})\log(\frac{1-\theta_k}{2}),\nonumber\\
\label{eq35}
\end{eqnarray}
with
\begin{eqnarray}
\theta_k=\sqrt{\sum_{j=1}^{3}q^{2}_{kj}}.
\label{eq36}
\end{eqnarray}
Therefore, the classical correlation for the spin pair is given by
\begin{small}
\begin{eqnarray}
\mathcal{C(}\rho)&=&\nonumber\\
&=&\max_{\{\Pi_i^B\}} \left(S(\rho^i)-\frac{S(\rho_0)+S(\rho_1)}{2}-c_3\cos\theta\frac{S(\rho_0)-S(\rho_1)}{2} \right).
\label{eq37}
\end{eqnarray}
\end{small}
 In general, $\mathcal{C(}\rho)$ has to be numerically evaluated by optimizing
over the angles $\theta$ and $\phi$. Once the classical correlation is obtained,
insertion of Eq.~(\ref{eq28}) and Eq.~(\ref{eq37}) into Eq.~(\ref{eq9}) can be used to determine the quantum discord.
\begin{figure}
\includegraphics[width=1.0\columnwidth]{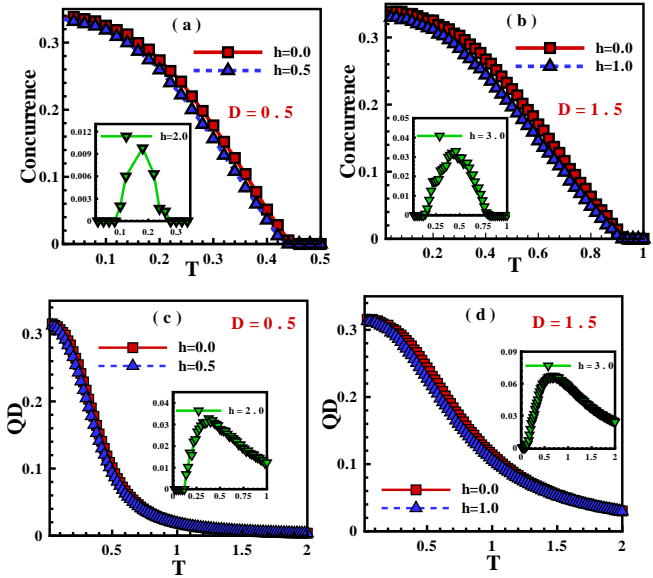}
 \caption{(Color online.) The concurrence ((a) and (b)) and QD ((c) and (d)) between NN spins as a
 function of the temperature for values of the LF less than quantum critical point and different
 values of the DM interaction. The inset shows the same figure for a value of the LF larger than $h_c$. }
 \label{Ent-QD-temperature}
\end{figure}

\section{Numerical results} \label{sec3}
Here we present our numerical results based on the theoretical formulation introduced in the previous sections.
. In Fig.~\ref{Ent-QD-temperature}, we have presented thermal behavior of the concurrence
(Fig.~\ref{Ent-QD-temperature}(a), (b)) and the QD (Fig.~\ref{Ent-QD-temperature}(c), (d)).
It can be  clearly seen that at zero temperature, NN spins are entangled in the quantum LL
phase ($h<h_c$). By increasing temperature both concurrence and QD start to decrease.
There is a field-independent critical temperature ($T_{c}$) point in which concurrence vanishes, whereas
QD shows an asymptotic behavior which is in agreement with results obtained by Bethe
ansatz approach\cite{Werlang10, Werlang11}.  In principle, at the mentioned critical temperature a part
of quantum correlations which contributes in the entanglement of formation will be destroyed
by classical thermal fluctuations. So at $T=T_{c}$ the concurrence is zero and one can derive the equation
\begin{eqnarray}
\gamma_{2}=-\sqrt{1-\sqrt{1-\gamma_{1}^{2}+\gamma_{1}^{4}}}.
\label{Tc}
\end{eqnarray}
By solving this equation, the critical temperature, $T_c$, will be obtained in whole range of parameters.
Note in the absence of the LF $\gamma_1=1/2$, at any value of the temperature and DM interaction,
thus the critical temperature can be recast as
\begin{eqnarray}
\gamma_{2}(T_{c}&,& h=0, D)=\frac{1-\sqrt{2}}{2} \\
&=&\frac{1}{2 \pi} \int_{-\pi}^{\pi} \cos(k) (\frac{a(k)}{\varepsilon(k)}) (\frac{1}{1+e^{\beta_c \varepsilon(k)}}-1/2) dk.\nonumber
\label{Tc}
\end{eqnarray}
 \begin{figure}
 \includegraphics[width=1.0\columnwidth]{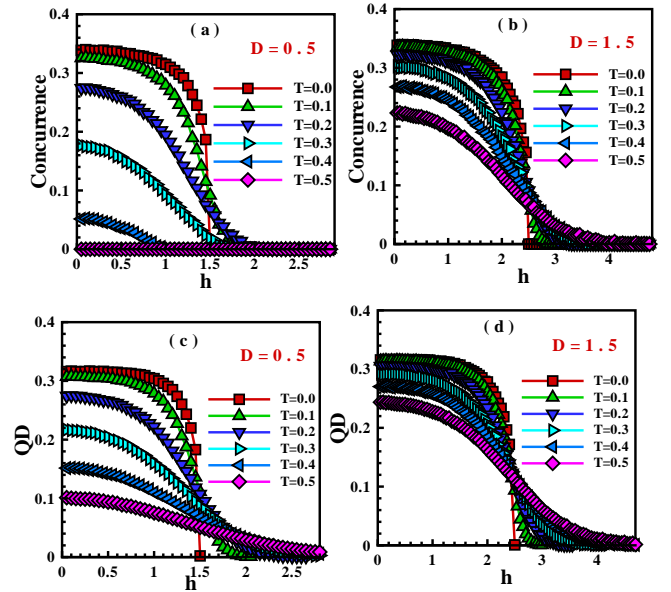}
 \caption{(Color online.) The concurrence ((a) and (b)) and QD ((c) and (d)) between NN
 spins as a function of the LF for different values of the temperature and DM interaction.   }\label{Ent-QD-h}
\end{figure}

Moreover,  for the LF more than the quantum critical point, NN spins are unentangled at zero temperature (see the inset of Fig.~\ref{Ent-QD-temperature}).
By increasing the temperature from zero, NN spins remain unentangled up to a critical  temperature, $T_{c_{1}}$. The existence of
this critical temperature can be related to the fact that the system is gapped in the saturated
ferromagnetic phase.  It shows that in the gapped saturated ferromagnetic phase, the quantum
correlations appear only at a finite value of the temperature proportional to the spin gap.
As soon as the temperature increases from $T_{c_{1}}$, both the concurrence and QD retrieve  and take
a maximum value. It be should noted that the amount of the $(QD)_{max}$ is almost three times bigger
than the maximum value of the concurrence. More increasing the temperature, the concurrence and QD
decrease and concurrence reaches to zero at a second critical temperature $T_{c_{2}}$, while QD
again shows asymptotic behavior.  This can be considered as an important difference between the QD and
concurrence. The existence of the second critical temperature shows that thermal fluctuations destroy
a part of quantum correlations related to the concurrence.

The effect of the LF on the quantum correlations is studied in Fig.~\ref{Ent-QD-h}. In this figure,
the concurrence (Fig.~\ref{Ent-QD-h}(a) and (b)) and the QD (Fig.~\ref{Ent-QD-h}(c) and (d)) are plotted
as a function of the LF. At zero temperature, NN spins are entangled in the absence of
the LF\cite{Jafari08}. When a LF applies the NN spins remain entangled up to the critical LF ($h_c$).
In the saturated ferromagnetic field the system becomes unentangled. It is worth to note that at zero temperature
the QD also show a critical field in which it is zero  in the saturated ferromagnetic phase.
At finite temperature, both concurrence and QD decrease with increasing the LF and vanishes in a
LF which depends on the temperature. Here we point out the DM interaction effects on quantum correlations.
Concurrence and QD show minor decreasing trend versus temperature for higher DM interaction.
It is obvious how the higher DM interaction in the system can persist over quantum fluctuations and
preserve the quantum correlations. This future is more profound in the concurrence. Indeed, in  Fig.~\ref{Ent-QD-h}
we have only chosen two DM interactions $D=0.5, 1.5$ to verify this fact. In what follows, we will focus on this issue in details.

\begin{figure}
 \includegraphics[width=1.0\columnwidth]{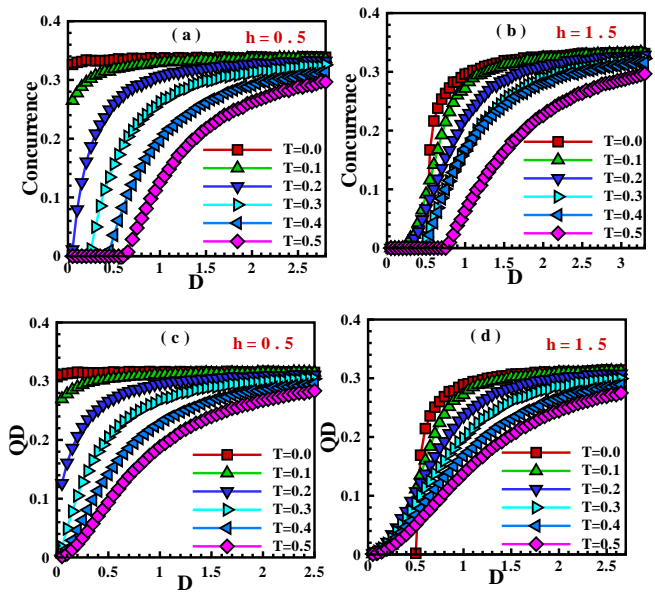}
  \caption{(Color online.) The concurrence ((a) and (b)) and QD ((c) and (d)) between NN spins as a
  function of the DM interaction  for different values of the temperature and LF.}
 \label{Ent-QD-D}
\end{figure}
  Fig.~\ref{Ent-QD-D} shows the behavior of the concurrence and the QD versus the DM interaction.
 When DM interaction is very small, the ground state of the system is in the Neel phase in the region $h<h_c$.
 In the Neel phase the NN spins are entangled \cite{Werlang10, Werlang11}. On the other hand, in the saturated
 ferromagnetic phase, $h>h_c$, NN spins are unentangled at zero temperature. It is obvious that by applying
 the DM interaction, concurrence and QD increase and reach to a saturation value. In addition, by
  increasing temperature all quantum correlations reduce, but only quantum correlations of the concurrence will
  be destroyed at a critical temperature. To this reason, a zero-plateau is observed in the curve of the concurrence
  (Fig.~\ref{Ent-QD-D}(a)). The existence of the zero-plateau in Fig.~\ref{Ent-QD-D}(b) is rooted to the quantum
   fluctuations  in the saturated ferromagnetic phase. The lake of zero-plateau in the curve of the QD confirms
   this fact that the quantum correlations of QD are very robust in comparison with concurrence.
\begin{figure}
\centerline{\psfig{file=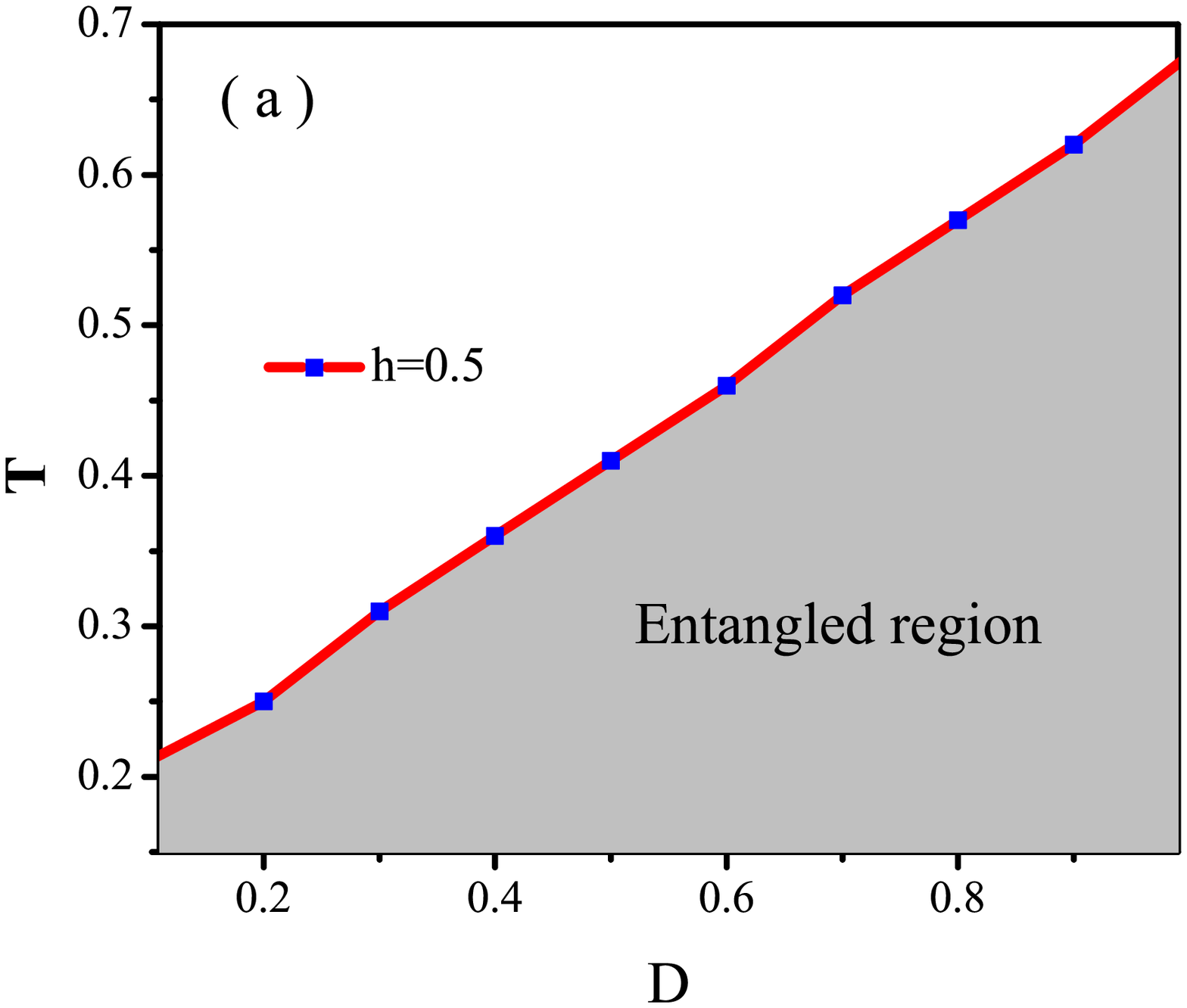,width=3.5in}}
\centerline{\psfig{file=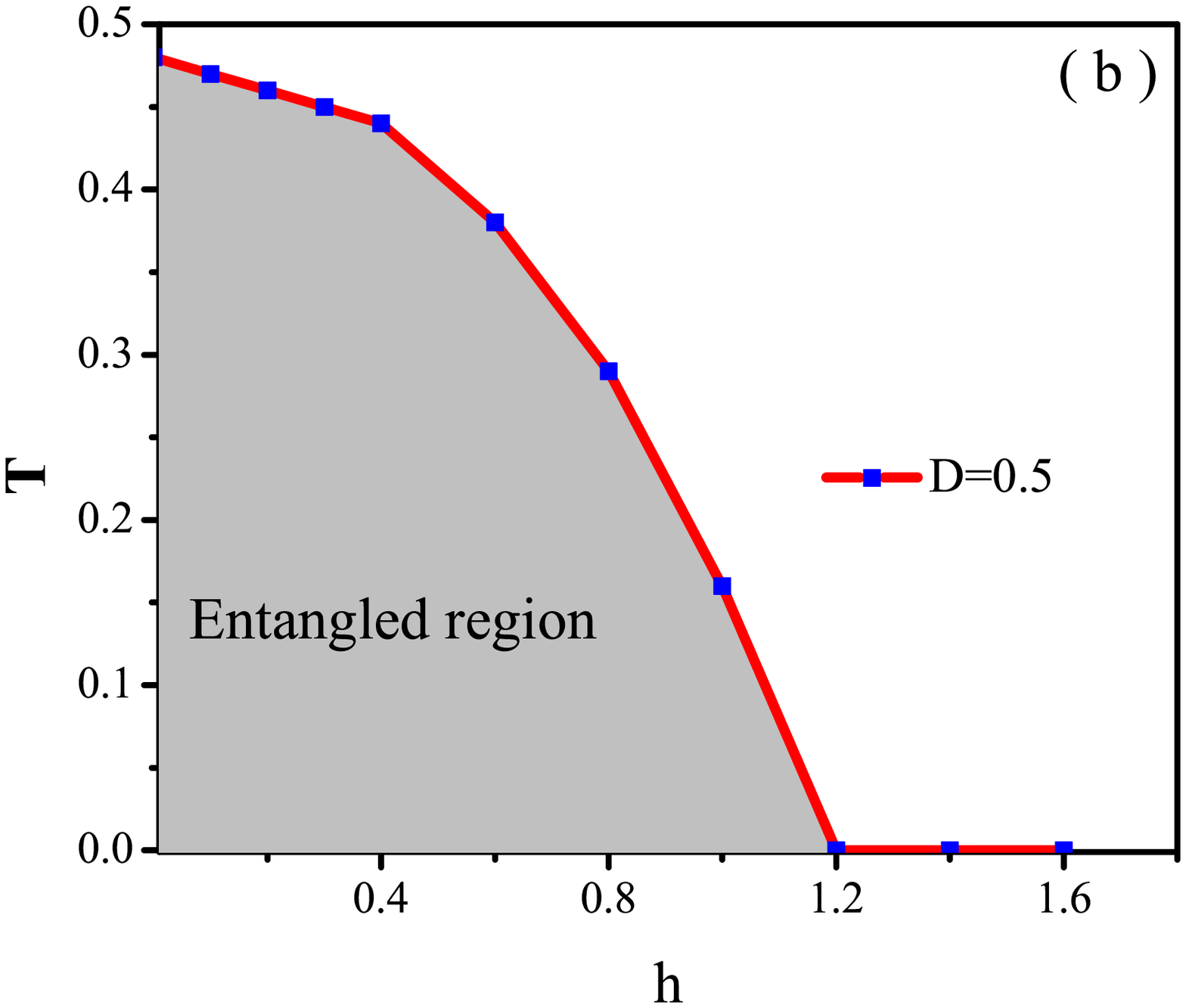,width=3.5in}}
 \caption{(Color online.) The parameter region of critical temperature $T$ and
 (a) the DM interaction ($h=0.5$), (b) the LF ($D=0.5$).}\label{witness-1}
\end{figure}
\section{Entanglement  witness} \label{sec3}
The realization that entanglement can also affect macroscopic properties of bulk solid-state systems,
has increased the interest in characterizations of entanglement in terms of macroscopic thermodynamical\cite{Ghosh03, Vedral03, Brukner04}
observables. The entanglement witness is called an observable which can distinguish between
entangled and separable states in the quantum physics\cite{Terhal00}. From an experimental point of view,
several methods for detection of entanglement using witness operators have been proposed \cite{Brukner04}.
 As a result of these studies, entanglement witnesses have been obtained in terms of expectation values of
 thermodynamic observables such as internal energy, magnetization and magnetic susceptibility.
As we have shown, the spin-orbit coupling introduces off diagonal terms in the Ising Hamiltonian.
In this case, results concerning Hamiltonian based on diagonal exchange interactions\cite{Brukner04, Toth05}
do not apply. In a recent work\cite{Wu05}, an appropriately generalized entanglement witness
is constructed where will be used in following.
Let $U=\langle H \rangle$ and $M=\sum_{n=1}^{N} S_{n}^{z}$ be the internal energy and the magnetization
along field respectively. Then Eq.~(\ref{eq15}) yields
\begin{eqnarray}
\frac{U+hM}{N}=\frac{1}{N}\sum_{n}\langle [D(S^{x}_{n}S^{y}_{n+1}-S^{y}_{n}S^{x}_{n+1})+J S^{z}_{n}S^{z}_{n+1}]\rangle .\nonumber\\
\end{eqnarray}
The right-hand of this equation is an entanglement witness. For any classical mixture of
products states (or separable state), the density matrix is written as
\begin{eqnarray}
\rho=\sum_{k}\omega_{k}\rho_{k}^{1} \otimes  \rho_{k}^{2}\otimes...\rho_{k}^{N},
\end{eqnarray}
where $\sum_{k}\omega_{k}=1$ and all $\omega_{k}\geq 0$. Using such density matrix, easily
one can find: $\langle S^{\alpha}_{n} S^{\beta}_{n+1}\rangle=\langle S^{\alpha}_{n}\rangle \langle S^{\beta}_{n+1} \rangle$ and $\sum_{\alpha=x, y, z} \langle S^{\alpha}_{n} \rangle^{2}\geq \frac{1}{4}$. Thus, the entanglement witness in the separable state has the upper bound as
\begin{eqnarray}
&|& \frac{1}{N}\sum_{n}\langle [D(S^{x}_{n}S^{y}_{n+1}-S^{y}_{n}S^{x}_{n+1})+JS^{z}_{n}S^{z}_{n+1}]\rangle |_{Max}=\nonumber \\
&|&ND+\frac{J-D}{N} \sum_{n}\langle S^{z}_{n}S^{z}_{n+1}]\rangle|.
\end{eqnarray}
Then the solid state system is in an entangled state if $W=\frac{|U+hM|}{ND}\geq \frac{1}{4}$.
Applying the fermionized operators, the entanglement witness is obtained as
\begin{eqnarray}
W=|\gamma_{1}^{2}-\gamma_{2}^{2}-\gamma_{1}+\gamma_{2}+\frac{1}{4}|.
\label{witness}
\end{eqnarray}
Using this equation we have determined the parameter regions where entanglement can be detected in
the solid state systems. Results are presented in Fig.~\ref{witness-1} (a) and (b). In the presence
of the magnetic field less than the quantum critical point (Fig.~\ref{witness-1}(a)), by adding DM
interaction, the critical temperature increases almost linearly in complete agreement with our results
on the concurrence and also Refs.\cite{Werlang10, Werlang11}. The effect of the LF on the critical
temperature is shown in Fig.~\ref{witness-1}(b). As it is seen the critical temperature decreases by
increasing the LF and will be zero for values of the LF $h(D=0.5)\geq 1.2$.
\section{CONCLUSION}\label{sec5 }
To summarize, we have studied the thermal quantum correlations in the 1D spin-1/2 Ising model with
added Dzyaloshinskii-Moriya (DM) interaction in the presence of a uniform longitudinal magnetic field(LF). First,
using the Jordan-Wigner transformation the model is transformed to a 1D  fermionized  model. Then, using the
mean-field approach the energy spectrum in an infinite chain is obtained.

By using the mean-field order parameters, we have determined the quantum discord (QD) and the concurrence between
 nearest neighbor (NN) spins at finite temperature. In principle, our approach is applicable to all cases contain the interacting fermions.
 A complete comparison between concurrence and the QD is done. At zero temperature, NN spins are entangled in the
 Neel phase ($h<h_c$) and unentangled in the saturated ferromagnetic phase ($h>h_c$). When system is in the Neel
  phase, the thermal fluctuations decrease all quantum correlations, but only destroy the quantum correlations of
   the concurrence. The quantum correlations of the QD exist even at finite temperatures which shows that the
   quantum correlations of QD are very stronger than the concurrence. On the other hand, when the ground state
   has the ferromagnetic long-range order,  there is not any quantum correlation at zero temperature. Increasing
    the temperature from zero, the NN spins remains unentangled up to a critical temperature which is rooted
    in the spin gap. More increasing the temperature, the concurrence and the QD regain  and take a maximum value.
     The amount of the $(QD)_{max}$ is almost three times larger than the maximum value of the concurrence.
     After this maximum, concurrence and QD decrease and concurrence reaches zero at a critical temperature while
     QD never goes zero.  The existence of the second critical temperature shows that thermal fluctuations
     will destroy a part of quantum correlations related to the concurrence but are not enough strong to
     destroy all quantum correlations.

Using macroscopic thermodynamic witnesses we have investigated quantum entanglement region in solids within
 our model. We have found that
 in the presence of the magnetic field less than the quantum critical point by adding DM interaction, the
 critical temperature increases almost linearly. The effect of the LF on the critical temperature is also
 shown that the critical temperature decreases with increasing the LF and will be zero for values of the LF $h(D=0.5)\geq 1.2$.
\vspace{0.3cm}


\end{document}